# Spatially resolved ultrafast transport current in GaAs photoswitches

L. Prechtel[1], S. Manus[2], D. Schuh[3], W. Wegscheider[4], and A.W. Holleitner[1,a]

[1]*Walter-Schottky Institut and Physik-Department, Technische Universität München,*

*Am Coulombwall 3, 85748 Garching, Germany*

[2]*Fakultät für Physik and Center for NanoScience (CeNS), Ludwig-Maximilians-Universität, Geschwister-Scholl-Platz 1, D-80539 München, Germany*

[3] *Institut für Experimentelle und Angewandte Physik, Universität Regensburg, D-93040 Regensburg, Germany*

[4] *Solid State Physics Laboratory, ETH Zurich, 8093 Zurich, Switzerland*

We apply a pump- and probe-scheme to coplanar stripline circuits to investigate the photocurrent response of GaAs photoswitches in time and space. We find a displacement current pulse, as has been reported earlier. A time-delayed second pulse is interpreted by a transport current. A time-of-flight analysis allows us to determine the velocity of the photogenerated charge carriers in the transport current. It exceeds the Fermi and the single-particle quantum velocities. This suggests that the excitation of a collective electron-hole plasma and not single charge-carriers dominates the ultrafast transport current in GaAs photoswitches.

---
[a] Author to whom correspondence should be addressed.
Electronic mail: holleitner@wsi.tum.de.

Commercial sources and detectors for terahertz electromagnetic radiation (THz-EMR) have lead to numerous applications in materials and life sciences. Nevertheless, the physical processes involved in THz-EMR generation are still subject of fundamental research. For instance, THz-EMR is generated by a non-linear wave conversion of a short laser pulse in an electro-optic medium,[1,2] by the photoelectric effect of a surface built-in electric field,[3] and by photo-Dember mechanisms.[4] Furthermore, hot carrier dynamics[5] and collective electron-hole plasma processes[6,7] can be probed by analyzing the emitted THz-EMR from a semiconductor test structure.

THz-EMR can also be generated in photoswitches (PS).[8] These are semiconductor slabs which are voltage-biased across two surface metal electrodes. The PS is typically excited by an optical femtosecond pulse. In turn, a short current response in the PS leads to THz-EMR. The optical illumination profile determines which optoelectronic phenomenon dominates the current response within the PS. A spatially uniform illumination of the PS gives rise to a transport current pulse which is limited by the lifetime of the optically excited charge carriers.[8,9] Since sub-picosecond carrier lifetimes can be achieved in a variety of materials (e.g. GaAs, Si, InGaAs), this kind of THz-EMR generation is widely exploited in scientific and technological applications. A non-uniform illumination leads to a dominating displacement current pulse whose duration can be shorter than the carrier lifetime.[10,11,12,13,14] While a theoretical description of both mechanisms has been developed,[14,15] the actual real-space motion of the transport current pulse has not been experimentally characterized so far.

Here, we report on space- and time-resolved photocurrent studies of a PS fabricated on low-temperature grown GaAs (LT-GaAs). The PS is part of a coplanar stripline circuit. An on-chip optoelectronic pump-probe scheme allows us to resolve the photocurrent reponse of the PS after a non-uniform illumination with micrometer and picosecond resolution. We find two photocurrent pulses in the time-resolved response of the PS. The first pulse is consistent with a displacement current pulse.[10,11,12,13,14] We interpret the second pulse to result from a transport current process. A time-delayed second pulse has been reported earlier, and its origin has been ascribed to spurious side-effects, such as THz-EMR reflections within the samples[16] as well as dispersive effects in the stripline, such as frequency dependent loss[17,18] and dispersion due to dielectric confinement.[19] Our spatially resolved experiments allow us to rule out these effects. By a time-of-flight analysis we can further determine the velocity of the photo-generated charge carriers of the transport current pulse. It exceeds the Fermi and the quantum velocity of single charge carriers. Hereby, we interpret the transport current pulse to stem from a collective electron-hole plasma excitation.[20]

Starting point is a heterostructure grown by molecular beam epitaxy on a GaAs substrate. In growth order, the layers are 350 μm GaAs, 2 nm AlGaAs, and 2 μm of LT-GaAs. After growth, the heterostructure is annealed at 600°C within a $O_2$ rich environment. A coplanar stripline (CPS) made out of Au with 200 nm height, 5 μm width, and a separation of 15 μm is fabricated by optical lithography.[21,22,23] All experiments are performed at a vacuum of ~$10^{-5}$ mbar and room temperature. First, the light of a titanium:sapphire laser is split into a pump- and a probe-pulse by an optical beam splitter. Both pulses are focused through an objective of a microscope onto the samples. The probe-pulse is focused at a position $x_0$ in between the two CPS. Hereby, the two metal strips of the CPS form a PS for the stripline circuit (Fig. 1). The pump-laser spot has a size of ~1.5 μm, and the laser pulse duration is ~160 fs. After optical pump-excitation, an electro-magnetic pulse starts to propagate along the CPS.[8] At a distance $d$, the field-probe of the sampling circuit is short-circuited by the probe-pulse for the duration of the lifetime of the photo-generated charge-carriers in the LT-GaAs at this position. The transient electric field of the CPS located at the field-probe during this time-period drives the current $I_{Sampling}$ in the sampling circuit (Fig. 1). Most importantly, varying the time-delay $\Delta t$ between the optical pump-pulse and the probe-pulse gives access to the time evolution of $I_{Sampling} = I_{Sampling}(\Delta t)$. The position of the pump-pulse $x_0$ in between the CPS is set by a scanning-mirror with a resulting spatial resolution of ~100 nm, while the position of the probe-pulse is kept constant throughout the experiments.

Fig. 2(a) depicts $I_{Sampling}$ as a function of $\Delta t$ for $d \approx 15$ μm and varying $x_0$. We observe two peaks of $I_{Sampling}$ [denoted by an open and a filled triangle in Fig. 2(a)]. For all $x_0$, we observe that the first peak occurs at the same $\Delta t$. Hereby, we interpret the first peak to result from a displacement current in the PS, which is consistent with the non-uniform illumination of a PS in stripline circuits.[10,11,12,13,14] The current-voltage characteristics of the CPS circuit without illumination demonstrate an ohmic transport between the Au and the LT-GaAs (data not shown). Therefore, we assume a linear bandstructure of the LT-GaAs in between the Au contacts.[24] Fig. 2(b) shows the resulting schematic of the bandstructure along $x_0$. An applied source-drain voltage $V_{SD}$ drops homogeneously across the gap between the CPS. The optical pump-pulse excites charge carriers in a region at $x_0$ with a diameter of ~1.5 µm [dashed lines in Fig. 2(b)]. Before illumination, the electric field $F = |\mathbf{F}| = V_{SD} / d$ is constant in between the two contacts [solid line in Fig. 2(c)]. The excitation laser generates a high local electrical conductivity in the vicinity of $x_0$.[14] Therefore, the electric field drops to zero at the position $x_0$ directly after laser excitation [dashed line in Fig. 2(c)], while it increases in the remaining area to sustain the applied bias $V_{SD}$.[12,13,14] The resulting displacement current density $j_D = \varepsilon_{GaAs}\varepsilon_0\, \partial \mathbf{F}/\partial t$ is coupled into the CPS and sampled at the field-probe ($\varepsilon_{GaAs}$ and $\varepsilon_0$ are the relative and vacuum permittivities).

The second peak in Fig. 2(a) is found to shift in $\Delta t$ with respect to the first peak as a function of $x_0$. The solid lines in Fig. 2(a) are least-squares fits with the sum of two Lorentzian curves at center positions $t_1$ and $t_2$. For clarity, only the Lorentzian curves fitting the second peaks are individually drawn as dashed lines. The difference $t_{12} = t_2 - t_1$ versus excitation position $x_0$ is plotted in Fig. 2(d) for excitation energies of $E_{Laser} = 1.51$ eV (filled circles) and 1.59 eV (open circles). We find values of $t_{12}$ in the range of 2.5 ps $\leq t_{12} \leq$ 7 ps. Most importantly, we find that for all $E_{Laser}$, the linear fits of the data intersect at the position of the grounded metal stripline ($x_0 \approx 0$ µm) for $t_{12} \approx 0$ ps with an absolute spatial error of 2 µm. The relative spatial error of the data in Fig. 2(d) is ~100 nm.[25]

The time-of-flight diagrams of Fig. 2(d) already allow us to rule out THz-EMR reflections within the samples to explain the second peak. A THz-EMR pulse, which is re-absorbed at the position of the field probe, would be expected at a time-delay of $\Delta t' = 2 \cdot w \, \varepsilon_{GaAs}^{1/2} / c \approx$ 8.4 ps, with $w$ = 350 µm the height of the substrate. A reflected THz-EMR pulse, generated by the displacement current, would therefore be expected at $\Delta t' \geq 8$ ps after the displacement pulse. At the same time, a THz reflection at the interface between the LT-GaAs and the GaAs layer of the heterostructure would be expected at $\Delta t \approx 0.1$ps. Therefore, THz reflections cannot account for the second peak of $I_{Sampling}$ for 2.5 ps $\leq t_{12} \leq$ 7 ps.

In order to rule out dispersion effects in the CPS,[16,17,18,19] we move the pump-spot close to the field-probe. Fig. 3(a) shows $I_{Sampling}$ as a function of $\Delta t$ for zero propagation length ($d \approx 0$ µm), at a fixed excitation position ($x_0 \approx 8$ µm), and for 1 V $\leq V_{SD} \leq$ 5 V. Again, we detect a second peak (denoted by a filled triangle) at about $t_{12}$ = 6 ps after the first peak (open triangle), which is consistent with the observations at $d \approx 15$ µm (Fig. 2). The amplitude of both the first and the second peak in Fig. 3(a) depend linearly on the applied bias $V_{SD}$ [Fig. 3(b)], as expected for displacement and transport currents. We would like to note that a variation of the excitation position $x_0$ is not feasible at zero propagation length $d \approx 0$ µm, because then, the generation of a "shoulder" superimposes the second peak at small $x_0$.[12]

We now turn again to the linear time-of-flight diagrams in 2(d). We would like to point out the following. The intersection of the linear fits at $x_0 \approx 0$ µm and $t_{12}$ = 0 ps cannot be explained by the reflection of THz-EMR within the samples nor by dispersive effects. However, a charge transport process perpendicular to the CPS can explain the time-of-flight graphs. The linear fits in Fig. 2(d) give propagation velocities of $v_{propagation} = \partial x_0 / \partial t_{12} = (1.3 \pm 0.2) \cdot 10^8$ cm/s and $v_{propagation} = (2.3 \pm 0.2) \cdot 10^8$ cm/s for $E_{Laser}$= 1.59 eV and 1.51 eV. These values exceed typical Fermi- and quantum velocities of single-particle charge excitations in GaAs at the utilized $E_{Laser}$. However, they are consistent with values for collective electron-hole plasma excitations within a GaAs PS.[20]

In ref. [20], charge plasma waves with $\sim 10^8$ cm/s have been reported, which oscillate perpendicular to the surface of the utilized heterostructures within the top LT-GaAs layer.

Therefore, we tentatively ascribe the second peaks in Figs. 2(a) and 3(a) to arise from the first initial transient oscillation of a plasma wave propagating parallel to the LT-GaAs surface. Experimentally, we find values for $v_{\text{propagation}}$ to be $\sim 10^8$ cm/s for 1.51 eV $\leq E_{\text{Laser}} \leq$ 1.72 eV with a small, non-systematic variation. In this energy range, charge carriers are not excited into the $L$- and $X$-side-valleys in GaAs.[5] In principle, the influence of the excitation power density must also be considered, since a phonon bottleneck can alter the decay dynamics of charge carriers in GaAs.[26] Therefore, it would be desirable to excite at an $E_{\text{Laser}}$ less than 35 meV above the band-gap of GaAs to avoid the excitation of optical phonons.[20] However, we find that our on-chip detection scheme is not sensitive enough to detect $I_{\text{Sampling}}$ for such an $E_{\text{Laser}}$.

Finally, we point out that reflections of the propagating electromagnetic pulse at the end of the CPS cannot explain the second photocurrent peak, since they are expected at $\Delta t > 350$ ps due to a total CPS length of ~45 mm. At the same time, our results suggest that the presented ultrafast transport current phenomena may occur in further GaAs based optoelectronic circuits comprising e.g. nanowires[21,22] and quantum wires.[23]

In summary, we study the spatial and temporal dependence of the photocurrent response of a photoswitch (PS) fabricated on LT-GaAs after a non-uniform illumination. We identify both a transport current and a displacement current pulse. The transport current pulse depends linearly on the lateral position between the two electrodes of the PS. A time-of-flight analysis allows us to extract the velocity of the photogenerated charge carriers. It is faster than the Fermi and the quantum velocity of single-particle excitations. We therefore conclude that a collective excitation of an electron-hole plasma and not single charge-carriers dominates the transport current on ultrafast timescales in GaAs photoswitches. The contribution of the displacement current does not depend on the excitation position within the PS, as has been reported earlier.

The authors thank J.P. Kotthaus and G. Abstreiter for support, and they gratefully acknowledge financial support by DFG-project HO 3324/4, the German excellence initiative via the "Nanosystems Initiative Munich" (NIM), and the "Center of NanoScience (CeNS)" in Munich. One of us (W.W.) acknowledges financial support from the Swiss Science Foundation (Schweizerischer Nationalfonds).

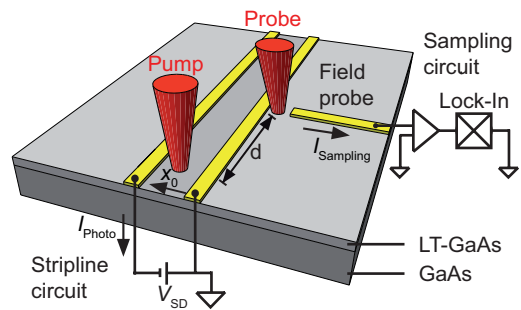

**FIG. 1**. Optoelectronic on-chip detection scheme. An optical pump-pulse is focused between two metal strips forming a coplanar stripline (CPS) at the position $x_0$. This gives rise to a photocurrent $I_{Photo}$ in the stripline circuit. At a distance *d*, the probe-pulse triggers the sampling circuit, which consists of a field probe, a current-voltage-amplifier, and a lock-in-amplifier. The (low-temperature grown LT) GaAs layer is depicted in dark (light) gray.

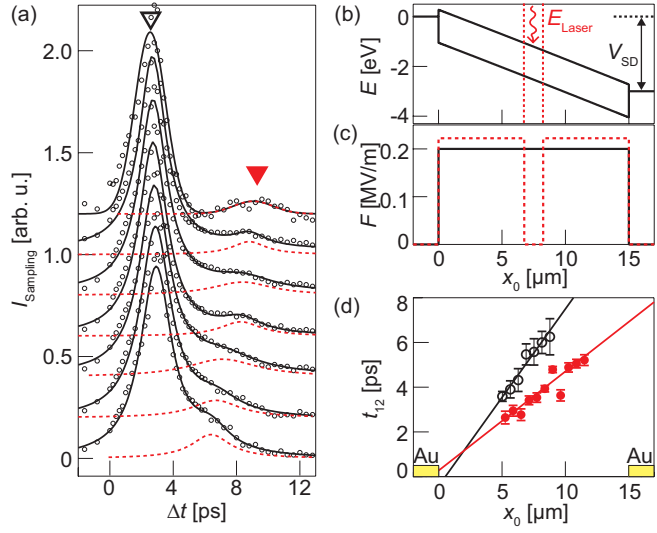

**FIG. 2.** (a) Time-resolved $I_{Sampling}$ for several excitation positions $x_0$, with 5.6 μm ≤ $x_0$ ≤ 9.4 μm in steps of 625 nm from bottom to top ($E_{Laser}$ = 1.59 eV, $P_{Laser}$ = 0.5 mW, $d ≈ 15$ μm, $V_{SD}$ = 3 V), with a first (open triangle) and a second peak (filled triangle). Graphs are off-set for clarity. Solid and dashed lines are fits. (b) Schematic band structure of GaAs along the $x_0$-direction at $V_{SD}$ = 3 V. The pump-laser excites charge carriers at position $x_0$. (c) Electric field $|\mathbf{F}|$ before (solid line) and directly after the pump-laser excitation (dashed line). (d) Relative time-delay $t_{12}$ between the two peaks in Figure 2(a) as a function of $x_0$ for $E_{Laser}$ =1.59 eV ($P_{Laser}$= 0.5 mW, open circles) and $E_{Laser}$ =1.51 eV ($P_{Laser}$= 880 μW, full circles). Lines are fits.

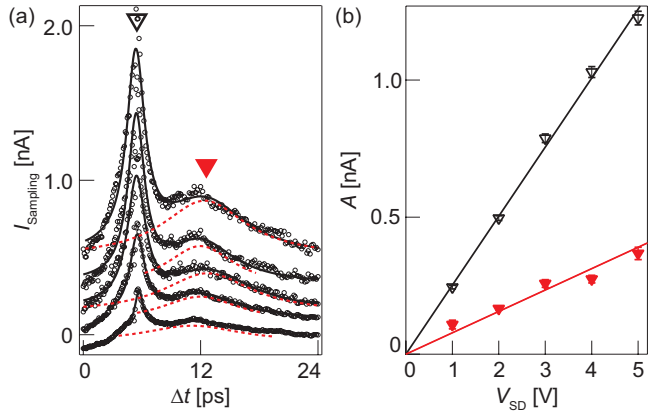

**FIG. 3.** (a) $I_{Sampling}$ as a function of $\Delta t$ for zero propagation distance ($d ≈ 0$ μm) and $V_{SD}$ = 1, 2, 3, 4, 5 V. Graphs are off-set for clarity ($E_{Photon}$ = 1.68 eV, $P_{Laser}$ = 1 mW, $x_0 ≈ 8$ μm). (b)

Amplitude of the first [open triangles] and second peak [filled triangles] as a function of $V_{SD}$. Dashed and solid lines are fits.